\pdfoutput=1
\documentclass[12pt,a4paper]{article} 
\usepackage{jcappub} 
\usepackage{amsmath}
\usepackage{amsfonts}
\usepackage{graphicx}
\usepackage{bm}
\usepackage{hhline}
\usepackage{subfigure}
\usepackage{multirow}
\usepackage{epstopdf}
\usepackage{epsfig}
\usepackage{varioref}
\usepackage{amssymb}
\usepackage{comment}

\def\ul{\underline}
\def\be{\begin{equation}}
\def\ee{\end{equation}}
\def\ba{\begin{eqnarray}}
\def\ea{\end{eqnarray}}
\def\bt{\begin{tabular}}
\def\et{\end{tabular}}
\def\btb{\begin{table}}
\def\etb{\end{table}}
\def\bfg{\begin{figure}}
\def\efg{\end{figure}}
\def\bc{\begin{center}}
\def\ec{\end{center}}
\def\bi{\begin{itemize}}
\def\ei{\end{itemize}}
\def\bl{\begin{list}}
\def\el{\end{list}}
\def\bn{\begin{enumerate}}
\def\en{\end{enumerate}}

\title{Observational constraints on K-inflation models}
\author{Sheng Li}
\author{and Andrew R. Liddle} 
\affiliation{Astronomy Centre, University of Sussex, Brighton BN1 9QH,
United Kingdom}

\emailAdd{sl277@sussex.ac.uk}
\emailAdd{a.liddle@sussex.ac.uk}
\abstract{
We extend the ModeCode software of Mortonson, Peiris and Easther \cite{ModeCode} to enable numerical computation of perturbations in K-inflation models, where the scalar field no longer has a canonical kinetic term. Focussing on models where the kinetic and potential terms can be separated into a sum, we compute slow-roll predictions for various models and use these to verify the numerical code. A Markov chain Monte Carlo analysis is then used to impose constraints from WMAP7 data on the addition of a term quadratic in the kinetic energy to the Lagrangian of simple chaotic inflation models. For a quadratic potential, the data do not discriminate against addition of such a term, while for a quartic ($\lambda \phi^4$) potential inclusion of such a term is actually favoured. Overall, constraints on such a term from present data are found to be extremely weak.
}

\begin{document}

\maketitle
\flushbottom

%%%%%%%%%%%%%%%%%%%%%%%%%%%%%%%%%%%%%%%%%%%%%%%%%%%%%%%%%%%%%%%%%%%%%%%

\section{Introduction}

Observations, especially including those from the Wilkinson Microwave Anisotropy Probe (WMAP) \cite{Dunkley:2008ie,Komatsu:2008hk,Komatsu:2010fb,Larson:2010gs}, are beginning to impose useful constraints on inflationary cosmologies. In particular, a number of papers have made detailed evaluations of constraints on the simplest inflation models, featuring a single canonically-normalized scalar field with unknown potential $V(\phi)$, see for example Refs.\ \cite{Dodelson:1997hr,Kinney:2002qn,wmappeiris,leli,Alabidi:2005qi,Peiris:2006ug,Kinney:2006qm,Ringeval:2007am,Kinney:2008wy,Hamann:2008pb,Adshead:2008vn,Agarwal:2008ah,Komatsu:2010fb}. 

Staying with the single-field paradigm, a more general scenario is available through the K-inflation paradigm. This retains minimal coupling of the scalar field to gravity, but allows the action to have an arbitrary dependence on the field's kinetic energy as well as on its value. This introduces new features, including a sound speed less than the speed of light which may enhance the non-gaussianity in the models. Our aim in this paper is to impose observational constraints on versions of this more general single-field scenario.

Our strategy is to modify the ModeCode program of Mortonson et al.\ \cite{ModeCode}, which solves the inflationary perturbation mode equations numerically and then interfaces to the CAMB \cite{CAMB} and CosmoMC \cite{CosmoMC} packages in order to compute the corresponding microwave anisotropies and compare to observational data. This entails a number of modifications to the way that ModeCode handles both the background (homogeneous) evolution equations and the perturbation equations. In this paper we will focus on the simplest case where the kinetic and potential terms remain sum-separable, and consider only simple forms for each. For the potential we will consider the simplest chaotic inflation models \cite{Linde:1981mu,Linde:1983gd}, based on quadratic and quartic potentials. For the kinetic term, we will consider simple monomial and polynomial forms, in particular investigating constraints on addition of a term quadratic in the kinetic energy to the normal canonical form. Future work will explore more complicated forms. Modification of ModeCode to consider specifically the action corresponding to DBI inflation (see Refs.~\cite{Sen:2002in,Sen:2002nu}) has already been carried out in Ref.~\cite{wmap-dbi}. Comparison of K-inflation models to five-year WMAP data using slow-roll methods has been made in Ref.~\cite{LMR}.

\section{The K-inflation model}

The K-inflation model \cite{KINFLATION,Kinfperts} features a single scalar field with the action
\begin{equation}
S = \int \sqrt{-g} \; p(\phi,X) d^4x \,,
\end{equation}
where $\phi$ is the field value and $X \equiv (1/2) \partial_\mu \phi \partial^\mu \phi$. The function $p$ will play the role of the pressure. A canonical scalar field has $p(\phi,X) = X - V(\phi)$ where $V(\phi)$ is the potential. In this paper we will focus on models where the kinetic and potential terms can be written as a sum:
\begin{equation}\label{larg}
p(\phi,X) = K(X) - V(\phi) \,,
\end{equation}
where $K(X)$ and $V(\phi)$ are both arbitrary functions to be determined from data.

Given the Lagrangian $p(X,\phi)$ of the considered model, we can obtain observable consequences by the following approach, closely following Garriga and Mukhanov \cite{Kinfperts}.

\subsection{Background field equations and sound speed}

We assume the usual Einstein equations and a spatially-flat Robertson--Walker metric with scale factor $a$ and Hubble parameter $H=\dot{a}/a$. We use the reduced Planck mass, defined by $M_{\rm Pl}^2 = (8\pi G)^{-1}$, to denote the strength of gravity throughout (with $c=\hbar=1$ as usual).

First, without needing to consider gravity, we can obtain a relationship between the density $\rho$ of the universe and its pressure $p$,
\begin{equation}\label{cont}
\dot\rho = -3H(\rho + p) \,,
\end{equation}
which is known as the continuity equation, in which the energy--momentum tensor is characterised by the pressure $p(X,\phi)$ and the density 
\begin{equation}
\rho(X,\phi) = 2X\frac{dp}{dX}-p \label{krho}
\end{equation}
of this universe. 

Second, now invoking a theory of gravity in the form of general relativity, we have the Friedmann equation for flat cosmologies
\begin{equation}\label{feq}
H^{2} = \frac{\rho}{3M_{\rm Pl}^{2}} \,.
\end{equation}
Taking Eq.~(\ref{cont}) together with Eq.~(\ref{feq}), we can therefore study the evolution of the scale factor $a$ and the field variable $\phi$.

One important quantity, called the `sound speed', describes the properties of the $\phi$ field. Regarding the field as a fluid system, we can introduce $c_{\rm s}^{2}$ as
\begin{equation}\label{defcs}
c_{\rm s}^{2} = \frac{p_{_{,X}}}{\rho_{_{,X}}} = \frac{p_{_{,X}}}{2Xp_{_{,XX}} + p_{_{,X}}}\,, 
\end{equation}
where the comma denotes the partial derivative with respect to $X$.

\subsection{Perturbation mode equations}

The background state for the field $\phi$ is given by Eqs.~(\ref{cont}) and (\ref{feq}). However, confrontation of inflationary models with data requires us to evaluate the perturbations they predict. We will just quote the two significant equations, first derived by Garriga and Mukhanov \cite{Kinfperts}. The scalar and tensor perturbations are described by quantities $v$ and $u$ whose Fourier components in the longitudinal gauge satisfy
\begin{eqnarray}\label{modes}
\frac{d^2 v_{k}}{d\tau^2} + \left(c_{\rm s}^{2}k^{2} - \frac{d^2 z/d\tau^2}{z}\right)v_{k} &=& 0 \label{scaleq} \\
\frac{d^2 u_{k}}{d\tau^2} + \left(k^{2} - \frac{d^2 a/d\tau^2}{a}\right)u_{k} &=& 0 \label{tenseq}
\end{eqnarray}
respectively. Here, $\tau$ is the conformal time, subscript $k$ denotes the momentum space, and the variable $k$ relates to the comoving scale by $\lambda = 2\pi/k$. The curvature perturbation $\zeta$ is related to $v$ by $\zeta = v/z $, while in a flat universe the background variable $z$ can be expressed as 
\begin{equation}
z^{2} = \frac{a^{2}(\rho + p)}{c_{\rm s}^{2}H^{2}} \,.
\end{equation}

In momentum space, the canonical quantisation modes $v_{k}$ and $u_{k}$ give two different classes of perturbations, the scalar perturbations measured by $v_{k}$ and the tensor (gravitational wave) perturbations by $u_{k}$. From Eq.~(\ref{scaleq}), we see that $c_{\rm s}/H$ plays the role of the `sound horizon', which the $k$ mode leaves by satisfying $c_{\rm s}k = aH$. The power spectrum of the scalar mode can be expressed by
\begin{equation}
\mathcal P_{\rm s}(k) = \frac{k^{3}}{2\pi^{2}}\left|\frac{v_{k}}{z}\right|^{2} \,,
\end{equation}
while in Eq.~(\ref{tenseq}) the tensor mode leaves the usual horizon by satisfying $k = aH$, with spectrum
\begin{equation}
\mathcal P_{\rm t}(k) = \frac{k^{3}}{2\pi^{2}}\left|\frac{u_{k}}{a}\right|^{2} \,.
\end{equation}

\subsection{Power spectra and observables}

Solving Eq.~(\ref{scaleq}) in the slow-roll approximation gives the expression \cite{Kinfperts}:
\be
\mathcal P_{\rm s} = \left. \frac{1}{8\pi^{2}M_{\rm Pl}^{2}} \, \frac{H^{2}}{\epsilon c_{\rm s}} \right|_{c_{\rm s}k=aH} \label{ps} \,,
\ee
The tensor mode Eq.~(\ref{tenseq}) has its usual power spectrum
\be
\mathcal P_{\rm t} = \left.  \frac{2}{\pi^{2}M_{\rm Pl}^{2}}H^{2}\right|_{k=aH} \label{pt}\,.
\ee

Having given the mode equations and their spectra, we can study the two most interesting variables, the spectral index $n_{\rm s}$ and the tensor--scalar ratio $r$. These two quantities can be constrained by observations, such as the existing WMAP data or the forthcoming Planck satellite results.  Their definitions are
\begin{eqnarray}
n_{\rm s} - 1 &\equiv& \frac{d\ln P_{\rm s}}{d\ln k} \simeq -2\epsilon - \tilde\eta- s \label{ns} \\
n_{\rm t} &\equiv& \frac{d\ln P_{\rm t}}{d\ln k} \simeq -2 \epsilon \label{nt}\\
r &\equiv& \frac{P_{\rm t}}{P_{\rm s}} = 16\epsilon c_{\rm s} \label{tsr} \,,
\end{eqnarray}
where the parameters $\epsilon\,,\tilde\eta\,,\delta\,,s$ are all small and defined as
\begin{eqnarray}
\epsilon = -\frac{d\ln H}{dN} \label{eps} \quad ; \quad
\tilde\eta = \frac{d\ln\epsilon}{dN} \label{teta} \quad ; \quad
\delta = -\frac{d\ln \dot\phi}{dN} \label{delt} \quad ; \quad
s = \frac{d\ln c_{\rm s}}{dN} \label{sounds} \,,
\end{eqnarray}
through $dN = Hdt$, and we additionally give the tensor spectral index $n_{\rm t}$. Higher-order versions of these expressions, which we do not use here, have been obtained using the uniform approximation \cite{LMR2}.

All these parameters are calculated at the time when mode $k$ leaves its individual horizon. For the scalar mode $k$ takes the value of $aH/c_{\rm s}$, while for tensor mode $k$ takes $aH$. As a result, the relation of $d/dN$ to $d/d\ln k$ for the scalars is
\begin{equation}\label{corr}
\frac{d\ln k}{dN} = 1 - \epsilon - s\,.
\end{equation}
Therefore Eq.~(\ref{ns}) and Eq.~(\ref{nt}) are then actually divided by Eq.~(\ref{corr}), whereas $s$ must be taken as zero when adjusting Eq.~(\ref{nt}) as the sound speed does not enter its horizon-crossing expression.

Note that $\tilde\eta$ in Eq.~(\ref{teta}) is implicitly a function of the usual slow-roll parameters $\epsilon=(M_{\rm Pl}^{2}/2)(V'/V)^{2}$ and $\eta = M_{\rm Pl}^{2} V''/V$ and other smaller parameters, such as $\delta$.\footnote{The full expression is 
\[ 
\tilde\eta = 2\epsilon - 2\left(1+\frac{Xp_{_{,XX}}}{p_{_{,X}}}\right) \delta + \frac{p_{_{,X\phi}}}{p_{_{,X}}} \phi_{_{,N}}
\]
given pressure $p = p(X,\phi)$, where ${,N}$ indicates the derivative with respect to $N$. In our current consideration, the pressure has separable $X$ and $\phi$, so the last term vanishes. The factor of $\delta$ in the second term can be treated as $-(1+\theta)$, where $\theta = 1/c_{\rm s}^{2}$. Therefore $\tilde\eta = 2\epsilon - (1+\theta)\delta$.} Here and throughout primes are derivatives with respect to $\phi$. We will discuss its particular expression in later sections when investigating specific models. 

\section{Slow-roll predictions}

\label{sec:sr}

In this section, we use the slow-roll approximation to compute the spectral index $n_{\rm s}$ and tensor-to-scalar ratio $r$ for various models. These results are of interest in their own right as they give an indication of the properties of models that will be able to fit the data. Additionally, we will be able to use them to verify that our modifications to ModeCode have been implemented successfully.

\subsection{General prediction without specifying a potential}

We study models where the Lagrangian takes the form
\begin{equation}
p(\phi,X) = K_{n+1}X^{n} - V(\phi) \label{lag}\,,
\end{equation}
where $K_{n+1}$ are constants and $n$ takes integer values.\footnote{With only a single kinetic term, the coefficient $K_{n+1}$ could be removed by rescaling $\phi$, adjusting the potential, but for later comparison with cases with more than one term we keep it explicit.} Under this kind of action, we can derive the field equation for the scalar field $\phi$ from Eq.~(\ref{cont}) as
\be
\dot{X}\rho_{_{,X}} + \dot\phi\rho_{_{,\phi}} = -6nK_{n+1}HX^{n} \label{eom-all} \,,
\ee
where the subscript $,\phi$ is the derivative with respect to the field.

We now discuss what the models predict under the slow-roll scheme. Therefore by the usual consideration, we assume the density of the universe is dominated by the scalar potential, $H^{2} \propto \rho \simeq V$ and replace all appearances of the Hubble parameter $H^{2}$ by $V$. In the field equation Eq.~(\ref{eom-all}), we take the first (acceleration) term on the left-hand side to be much less than the second term. Hence it simplifies to
\be
V^{\prime}\dot\phi \simeq -6nK_{n+1}HX^{n} \label{eom-sl}\,.
\ee
Within the slow-roll assumption, we can obtain  descriptions of the observable quantities by means of some parameters. From Eq.~(\ref{eps}) we can identify $\epsilon$ as $\epsilon_{_{V}}$ under the slow-roll assumption, where
\be
\epsilon_{_{V}} = -\frac{1}{2} \frac{V^{\prime}}{V} \frac{\dot\phi}{H} \label{sl5} \,.
\ee
With $\ddot\phi \ll \dot\phi$ under the slow-roll assumption,  for later discussion we give the second parameter $\eta_{_{V}}$ which relates to the first and second-order derivatives of the potential,
\be
\eta_{_{V}} = -\frac{V^{\prime\prime}}{V^{\prime}}\frac{\dot\phi}{H}\label{sl6} \,.
\ee
By the approximation above, we can find a simple relation between these parameters,
\be
\frac{\eta_{_V}}{\epsilon_{_{V}}} = 2\frac{V V^{\prime\prime}}{V^{\prime 2}}\,,
\ee
and also we can find the relation
\be
\tilde\eta = 3\epsilon_{_{V}} - \delta - \eta_{_{V}} \label{releta}\,.
\ee
From Eq.~(\ref{eom-sl}), for later evaluation we can write the ratio $\dot\phi/H$ as
\be
\frac{\dot\phi}{H} = -\alpha(n) \left( \frac{V^{\prime}}{V^{n}} \right)^{1/(2n-1)}  \label{hdotphi} 
\ee
where 
\be
\alpha(n) = \left( \frac{6^{n-1}}{nK_{n+1}} {M_{\rm Pl}^{2n}}\right)^{1/(2n-1)} \label{alp-sl}\,.
\ee
Then 
\be
\epsilon_{_{V}} = \frac{1}{2}\alpha(n)\left(\frac{V^{\prime 2n}}{V^{3n-1}}\right)^{1/(2n-1)} \quad; \quad
\eta_{_{V}} = \alpha(n)\left(\frac{V^{\prime\prime (2n-1)}}{V^{n}V^{\prime 2n-2}}\right)^{1/(2n-1)} \,.
\ee

Now we would like to apply the slow-roll approximation to evaluate the most important physical observables. The first is the scalar power spectrum $\mathcal P_{\rm s}$,
\be
\mathcal P_{\rm s} \propto  \frac{2}{\alpha(n) c_{\rm s}} \left(\frac{V^{5n-2}}{V^{\prime 2n}}\right)^{1/(2n-1)} \label{ps-sl} \,, 
\ee
In this case, we can write the spectral index via Eq.~(\ref{ps-sl}) as
\be
n_{\rm s} - 1 = \frac{1}{2n-1}\left[ 2n\eta_{_{V}} - 2(5n-2)\epsilon_{_{V}}\right] \label{index-sl} \,.
\ee
To verify that Eqs.~(\ref{ps-sl}) and (\ref{index-sl}) are correct for standard inflation, we just need to set $n=1$, and then we have the power spectrum 
\be
\mathcal P_{\rm s} = \frac{1}{12\pi^{2}M_{\rm Pl}^{2}} \frac{K_{2}V}{2\epsilon_{_{V}}}\,,
\ee 
and its corresponding spectral index is $n_{\rm s} - 1 = 2\eta_{_{V}} - 6\epsilon_{_{V}}$. As a prediction of  Eq.~(\ref{index-sl}),  for the simplest non-canonical inflation, in which the kinetic term has the form $X^{2}$, we can obtain the power spectrum 
\be
\mathcal P_{\rm s} = \frac{1}{12\pi^{2}M_{\rm Pl}^{4}} \left(\frac{K_{3}V^{8}}{3M_{\rm Pl}^{4}V^{\prime 4}}\right)^{1/3} \,,
\ee
and its spectral index $n_{\rm s} - 1 = ( 4\eta_{_{V}} - 16\epsilon_{_{V}})/3$.

\subsection{Predictions for specific models through e-folding $N$}

Thus far we have obtained the formulae for the scalar power spectrum and its spectral index without specifying a particular potential type. In this subsection, we continue the discussion within a class of potential $V(\phi) = A \phi^{m}$ in the Lagrangian Eq.~(\ref{lag}), where $A$ denotes the normalisation parameter, and $m$ takes integer values.\footnote{Following footnote 2, for this form of potential a rescaling of $\phi$ to eliminate $K_{n+1}$ in the single-term kinetic case would simply renormalize $A$, which anyway is to be fixed by the density perturbation amplitude. Hence there is a perfect degeneracy between $K_{n+1}$ and $A$ and data can only fix a combination of them.}

The e-folding number $N$, which measures how much inflation took place, is
\be
N = -\int H \,dt \label{efold} \,.
\ee
After observable scales cross the horizon we have $N \sim 50$ until inflation ends. First we evaluate the time variation $H\,dt$ by the time-shift carried by the scalar field $\phi$ and finally transferred from the gradient change in the potential $V(\phi)$
\be
H\,dt = \frac{H}{\dot\phi}d\phi = \frac{H}{\dot\phi} \frac{1}{ V^{\prime}}dV \,.
\ee
So from the definition of $N$, and taking the above relation along with Eq.~(\ref{hdotphi}), we can derive the corresponding relation for $N$ in terms of the scalar potential $V(\phi)$ and its derivatives, rather than with $\phi$ itself,
\be
N = \int^{\rm e}_{\rm i} \frac{H}{\dot\phi} \frac{1}{V^{\prime}} dV \label{npotential} \,.
\ee
Under the assumption of slow-roll, by means of the slow-varying parameters, say Eqs.~(\ref{sl5}) and (\ref{sl6}), we can write Eq.~(\ref{npotential}) as
\be
N =  -\frac{1}{2}\int^{{\rm e}}_{{\rm i}} \frac{1}{\epsilon}\frac{dV}{V} \label{nepsv}\,.
\ee
Here $\epsilon$ (neglecting the subscript) is a function of the potential and its derivatives as well, which have already been denoted by Eqs.~(\ref{sl5}) and (\ref{sl6}) in the slow-roll approximation. Connecting $N$ with these two slowly-varying parameters leads to the expression for spectral index $n_{\rm s}$.

By taking potential $V(\phi) = A \phi^{m}$, we obtain
\be
N = \frac{m}{\beta(n,m)} \frac{1}{\gamma(n,m)} V^{\beta(n,m)/m}  \label{npot} 
\ee
where
\be
\beta(n,m) = \frac{m(n-1)+2n}{2n-1} \quad ; \quad 
\gamma(n,m) =  \alpha(n) \left( mA^{1/m} \right)^{2n/(2n-1)}   \label{coec} \,.
\ee
Then, connecting Eqs.~(\ref{sl5}), (\ref{sl6}) and Eq.~(\ref{npot}), 
we can get the relation between the slow-roll parameters and $N$
\be
\epsilon_{_{V}} = \frac{m}{2\beta} \frac{1}{N} \quad ; \quad \eta_{_{V}} = \frac{m-1}{\beta} \frac{1}{N} \label{etan} \,. \ee
The final formula for the spectral index in Eq.~(\ref{index-sl}) is therefore 
\be
n_{\rm s} - 1 = - \mathcal{I}(n,m)\frac{1}N \label{nindex} \,,
\ee
where
\be
\mathcal{I}(n,m) =  \frac{m(3n-2) + 2n}{m(n-1)+2n} \label{indexi}\,.
\ee

We can separate the two dependencies in Eq.~(\ref{indexi}) so as to examine how the potential and kinetic energy terms contribute:
\be
\mathcal{I} = 1 + \frac{(2n-1)m}{ m (n-1) + 2n } = 1 + \frac{m}{\beta} \label{CI} \qquad \mbox{($n\,,m \ge 1$)} \,,
\ee
which indicates $\mathcal I > 1$. We can see, from the above expression, that the spectral index $n_{\rm s} - 1$ has a simple relation with the power of the potential in $N$, Eq.~(\ref{npot}). In particular, for the quadratic potential $m=2$, $\mathcal{I}$  is independent of $n$. Therefore, the inflation model driven by this potential will give a spectral index $n_{\rm s} = 0.96$ regardless of the power of $X$ in the kinetic term.
Further, in terms of $\mathcal I(n,m)$ in Eq.~(\ref{CI}), the scalar power spectrum Eq.~(\ref{ps}) can be written as
\be
\mathcal P_{\rm s} = \frac{1}{12\pi^{2}M_{\rm Pl}^{4}} \frac{1}{c_{\rm s}} {\gamma}^{\mathcal I - 1} \left( \frac{N}{\mathcal I - 1} \right)^{\mathcal I} \label{ps-ci}\,.
\ee

The class of Lagrangian, Eq.~(\ref{lag}), has sound speed $c_{\rm s} = 1/\sqrt{2n-1}$ which is independent  of $m$. As for Eq.~(\ref{ps-ci}), given a known e-folding number $N$ after inflation, the scalar power spectrum can be determined directly using the parameter set $(n,m)$, which is the input argument to the two functions $\gamma(n,m)$ and $\mathcal I(n,m)$, while for the scalar spectral index we can just use one function $\mathcal I(n,m)$. Example values for the functions $\alpha(n)$, $\beta(n,m)$ and $\gamma(n,m)$ with parameter set $(n,m)$ can be seen in Table~\ul{\ref{abg}}.

\btb[t]
\centering
\bt{|c|c|c|c|c|}
\hline
	&\multicolumn{2}{c|}{$n=1$}&\multicolumn{2}{c|}{$n=2$}\\ \cline{2-5}
	&$m=2$	&$m=4$	&$m=2$	&$m=4$	\\
	\hline
$\alpha(n)$ & $M_{\rm Pl}^{2}/K_{2}$ & $M_{\rm Pl}^{2}/K_{2}$ & $(3M_{\rm Pl}^{4}/K_{3})^{1/3}$ & $(3 M_{\rm Pl}^{4}/K_{3})^{1/3}$ \\ \hline
$\beta(n,m)$ & 2 & 2 & 2 & 8/3\\ \hline
$\gamma(n,m)$ & $4\alpha(1)A$ & $16\alpha(1)A^{1/2}$ & $2^{4/3}\alpha(2)A^{2/3}$ & $2^{8/3}\alpha(2)A^{1/3}$\\ \hline
$\mathcal I(n,m)$ & 2 & 3 & 2 & $5/2$ \\ \hline \hline
$\mathcal P_{\rm s}$ & $\gamma_{(1,2)}N^{2}$ & $\gamma^{2}_{(1,4)} (N/2)^{3}$ & $\sqrt{3}\gamma_{(2,2)} N^{2}$ & $ \sqrt{3}\gamma^{3/2}_{(2,4)} (2N/3)^{5/2}$ \\ \hline
$n_{\rm s}\big|_{_{N = 50}}$ & 0.96 & 0.94 & 0.96 & 0.95  \\ \hline
%$r \big|_{_{N = 50}}$ & 0.16 & 0.32 & $0.16/\sqrt{3}$ & $0.08*\sqrt{3}$  \\ \hline
$r \big|_{_{N = 50}}$ & 0.16 & 0.32 & $0.092$ & $0.14$  \\ \hline
\et
\vskip .1in
\caption{Functions and observables for standard inflation and the simplest non-canonical inflation (NCI)  model $K(X) = K_{3}X^{2}$. $\mathcal P_{\rm s}$ is in units of $(12\pi^{2}M_{\rm Pl}^{4})$ and $N$ is taken as 50.}\label{abg}
\etb
%%%%%%%%%%%%%%%%%%%%%%%%%%%%%
%%%%%%%%%%%%%%%%%%%%%%%%%%%%%

\bfg
\centering
\includegraphics[width=0.8 \textwidth]{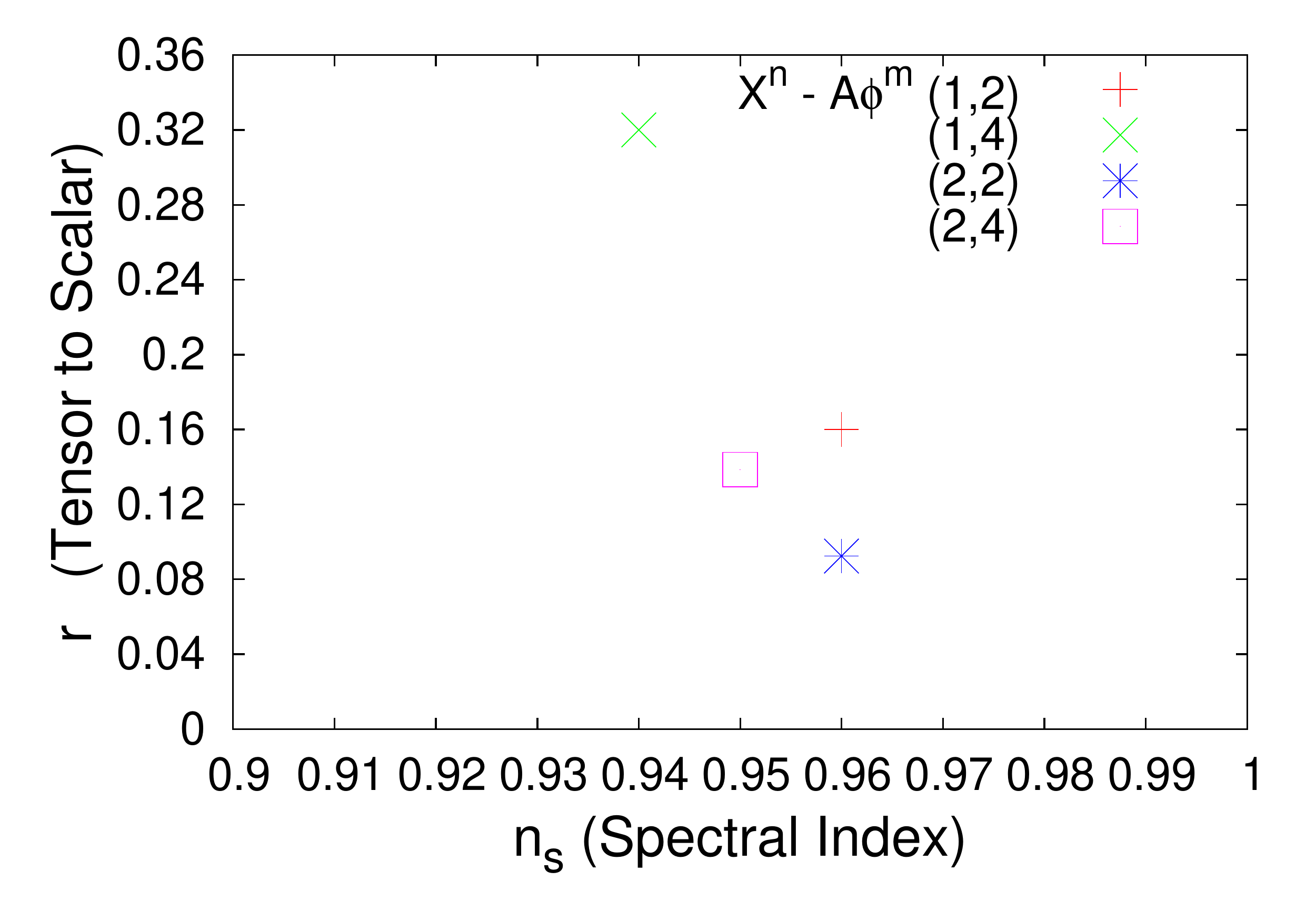}
\caption{Slow-roll predictions for the tensor-to-scalar ratio $r$ and the scalar spectral index $n_{\rm s}$ in standard inflation and the simplest NCI model where $n=2$ in Eq.~(\ref{lag}).}{\label{ratio-ns}}
\efg
%%%%%%%%%%%%%
%%%%%%%%%%%%%

From Eqs.~(\ref{tsr}), (\ref{etan}) and (\ref{nindex}), we can obtain an expression relating the tensor-to-scalar ratio $r$ and spectral index $n_{\rm s}$
\be  \label{rns-sl}
r = 8 \frac{\sqrt{2n-1}\, m}{m(3n-2)+2n}\left( 1-n_{\rm s} \right) \,.
\ee
We list the observables in Table~\ul{\ref{abg}}, and example values for $n_{\rm s}$ and $r$ can be seen in Figure~\ul{\ref{ratio-ns}}.

For canonical inflation, the known relation $n_{\rm s} - 1 = - (m+2)/2N$ is recovered by setting $n=1$. For $n=2$, we find 
\be\label{ns-m-sl}
n_{\rm s} - 1 = -\frac{4(m+1)}{m+4} \frac{1}{N} \,. 
\ee
Figure~\ul{\ref{fig:ns-nm-sl}} shows the spectral index as a function of $m$ for several $n$ values; when $n>1$ the spectral index asymptotes to $1-4/N$ in the limit of large potential power-law $m$, unlike the canonical case where $1-n_{\rm s}$ grows linearly with $m$ and can be large.

%%%%%%%%%%%
%%%%%%%%%%%
\bfg
\centering
\includegraphics[width=0.8 \textwidth]{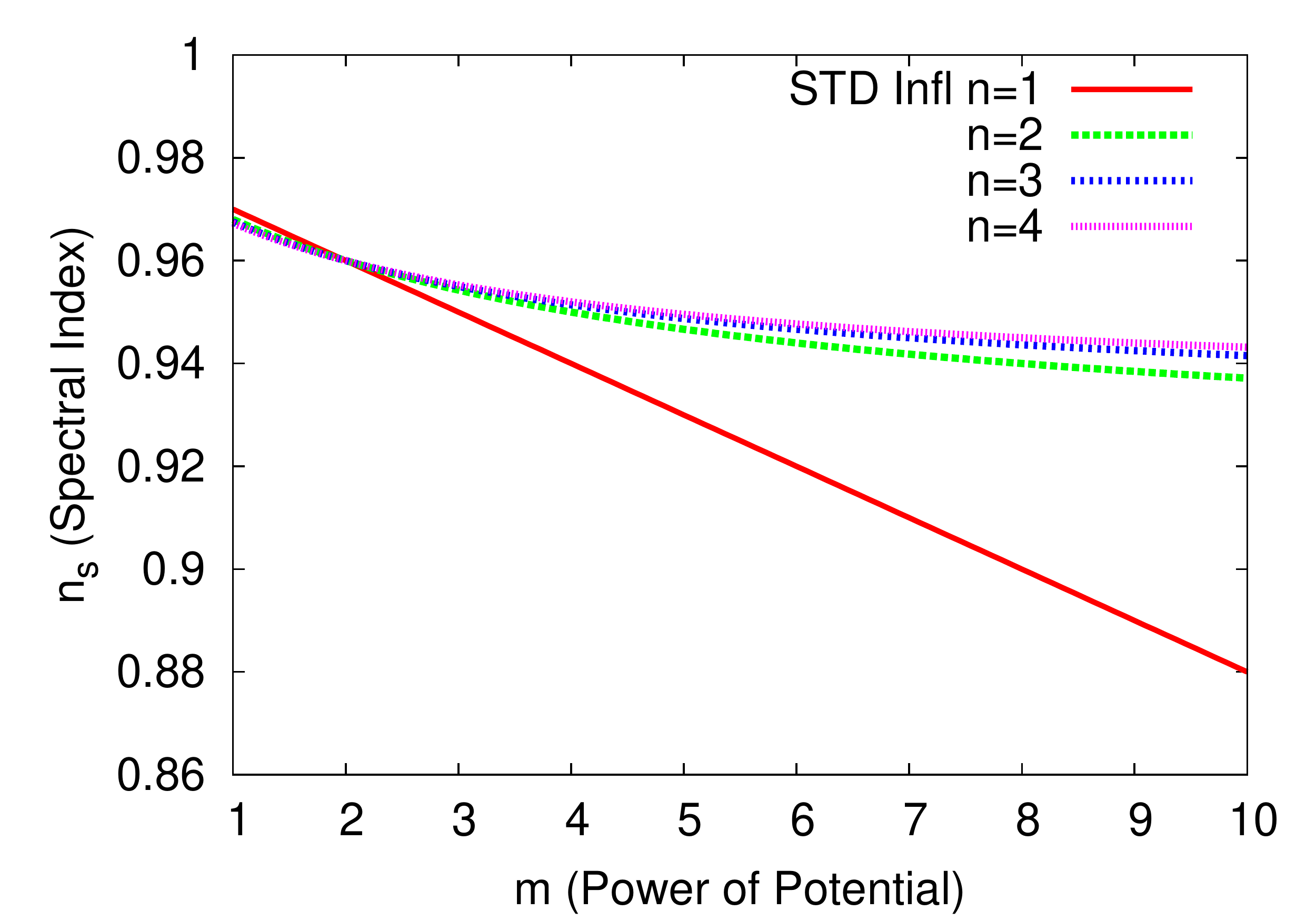}
\caption{Slow-roll predictions for the spectral index with different kinetic power-law $n$ and potential power-law $m$. This plot assumes the pivot corresponds to $N=50$. }\label{fig:ns-nm-sl}
\efg
%%%%%%%%%%%
%%%%%%%%%%%

\section{ModeCode for K-inflation}

For single-field canonically-normalized inflation models there are many numerical tools which can calculate the primordial power spectra, as well as other characteristics such as the bispectrum and trispectrum of non-gaussianities. Also many of these tools have been interfaced with MCMC codes such as CosmoMC so as to explore the likelihood and carry out parameter estimation. ModeCode \cite{ModeCode} is an example of such a programme, which had also recently been interfaced to the MultiNest model selection code \cite{mc2,mc3}. Other codes to numerically solve the inflationary mode equations have been described in Refs.\ \cite{gl96,Martin:2006rs, Lesgourgues:2007gp,Lesgourgues:2007aa,Finelli:2009bs, Martin:2010hh}.

To study non-canonical models, modifications must of course be made to the code, generalizing both the background and perturbation equations. An example already in the literature is an enhancement to consider the Lagrangian describing DBI inflation, made in Ref.~\cite{wmap-dbi}, which is motivated from string scenarios \cite{Sen:2002in,Sen:2002nu} and has been widely studied in the literature. In this paper we consider a different extension to non-canonical inflation models, at this stage restricted to models where the kinetic energy $K(X)$ and the potential $V(\phi)$ are sum-separable.

\subsection{Brief description of ModeCode}

ModeCode is characterised by free parameters describing the inflationary potential for canonical single-field inflation, and solves the inflationary mode equations numerically, bypassing the slow-roll approximation. It computes the cosmic microwave background angular power spectra and performs a likelihood analysis and parameter estimation by interfacing with CAMB \cite{CAMB} and CosmoMC \cite{CosmoMC}.

\subsection{Modifications needed for K-inflation}\label{kmcmod}

We implement ModeCode under our umbrella \textbf{Kinetic Module Companions (KMC)}. To perform our analysis for NCI models, we build up a full system for initialising the background equations, avoiding slow-roll or any other kind of approximation beyond linear perturbation theory. The parameters which can be explored and the methodology in KMC are as follows.
\bi
\item Parameters describing the form of the kinetic energy, for example a Taylor expansion of $K(X)$ about $X=0$ which we have currently implemented up to sixth order (though in the present paper we will only consider up to quadratic order).
\item Parameters describing the inflationary potential, for instance the potential can take a polynomial form or be a Taylor series.
\ei
We use eigenvalue methods to get the real solutions needed by the background equations, Eqs.\ (\ref{krho}) and (\ref{eom-sl}), as well as perturbed equations.
Equations to solve simultaneously for  $\phi_{,_{N}}$ and $H^{2}$ can be explicitly obtained.

\subsection{Comparison tests}

\subsubsection{Recovery of ModeCode results for $K(X) =X$}

Before running the extended functions, we check we can recover the results of Mode\-Code with our KMC system. The outputs that ModeCode generates are recovered either by setting our flag \textcolor{blue}{use\_kinetic=T} and setting $K(X)=X$, so that KMC will perform its intrinsic functions for models that can be executed by ModeCode, or just by switching off the KMC functional system by flag \textcolor{blue}{use\_kinetic=F}, leaving KMC to function as the normal ModeCode. We have confirmed that the ModeCode results are then precisely recovered under either method.

\bfg
\centering
\includegraphics[width=0.8\textwidth]{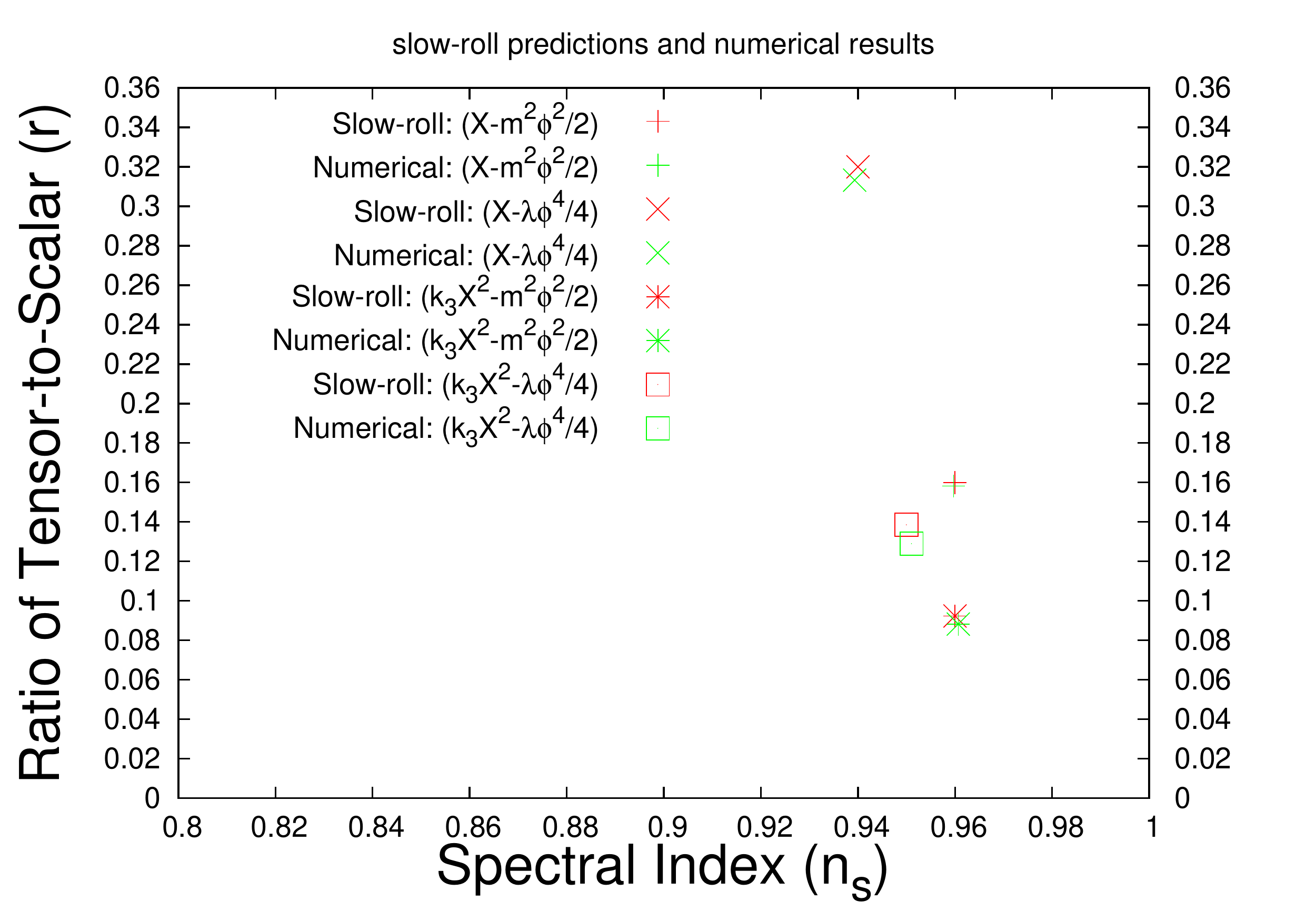}
\caption{Slow-roll predictions and numerical results for standard canonical inflation and the simplest NCI model ($K_{3}X^{2} - A\phi^{m}$), showing they are consistent. The matter power spectrum amplitude constrains the combination $AK_3^{m/4}$ (from the scaling argument of footnote 3) and the predictions for $n_{\rm s}$ and $r$ are independent of this. The value of $K_{3}$ does determine the e-folding value corresponding to observable scales, but this is not fixed by observations.}
\label{fig:sl-num}
\efg
%%%%%%%%%%%

\subsubsection{Recovery of slow-roll results}

Now we compare the slow-roll predictions of the previous section with the numerical results for two types of inflation model. One is the standard inflationary model, with canonical kinetic energy and Lagrangian $p(\phi,X) = X - V(\phi)$. The other is the simplest non-canonical inflation model, where the kinetic energy takes the form $X^{2}$ in the Lagrangian $p(\phi,X) = K_3 X^{2} - V(\phi)$. Figure~\ul{\ref{fig:sl-num}} shows that  the recovery of the slow-roll results is very accurate. It is not expected to be absolutely precise because the slow-roll approximation is not perfect, for instance leading to an offset in identification of the $N=50$ point as well as neglecting higher-order corrections to perturbation observables.
%%%%%%%%%%%

\section{Parameter explorations with MCMC methodology}
\subsection{Global settings and initial conditions}

We now proceed to explore the parameter space of a particular non-canonical type of inflationary model, by means of Markov Chain Monte Carlo (MCMC) methodology. We use the WMAP 7-year data version4 (``WMAP7''), and our simulations take 12 chains for each model. We set the pivot scale to $k_{\rm pivot}=0.05 \, {\rm Mpc}^{-1}$. We aim to select an initial field value $\rm{\phi_{\rm init}}$ which corresponds to 70 $e$-foldings from the end of inflation, estimated analytically by assuming a single power-law term dominates $K(X)$; if this approximation proves inaccurate it gets adjusted by the numerical code. We then must choose a consistent initial field velocity $\rm{\phi_{_{,N}{\rm init}}}$, as mentioned in Section~\ref{kmcmod}, a task which has been solved by eigenvalue methods in our \textcolor{blue}{KMC} numerical modules. 

Having developed the KMC code, our aim is to investigate what type of NCI models are supported by observational data. Using the MCMC method, we will perform a likelihood analysis and parameter exploration for some particular non-canonical inflation models.

\subsection{Choice of models}

In this article, for our numerical work we focus on a particular choice of kinetic term which adds a quadratic term in $X$ to the usual linear one. Investigation of more complex models will be made in future work. Hence our considered NCI model is $p(X,\phi) = K_{2}X + K_{3}X^{2} - V(\phi)$ with $K_3$ positive,\footnote{Negative $K_3$ appears possible in principle, provided $X$ is not too large, but gives models that can have phantom behaviour ($w <-1$) from an overall negative kinetic term, which can be expected to cause instabilities. We note also that the quadratic approximation to the DBI model features a positive coefficient.}   and we will additionally assume the potential to have a single polynomial term $V(\phi)=A\phi^{m}$ with $m=2$ or $4$, giving a large-field model. The field $\phi$ can always be rescaled to set $K_2 = 1$, and the new term with coefficient $K_3$ can be considered as the first correction term in a Taylor expansion of a general $K(X)$ that reduces to a canonical form in the limit $X \rightarrow 0$. While such a model is not particularly realistic, it has the benefit of simplicity and it is interesting to ask whether present data can say anything about the possible values of such a correction. Since in slow-roll inflation $X$ will be small, we can immediately anticipate that any constraint on $K_3$ will be very weak, allowing values much greater than one before this correction term could significantly modify the canonical term. 

Our free parameters are therefore the power of the potential, which we fix for each investigation, and the values of the amplitude of the potential and the coefficient $K_3$. Additionally, the value of $N_{\rm pivot}$ corresponding to $k_{\rm pivot}$ can vary as in the original ModeCode. The well-measured amplitude of perturbations will accurately fix a combination of these parameters.

\subsection{Overview of effects from additional kinetic terms}

\bfg[t]
\centering
\subfigure[plot for quadratic potential]{\label{fig:tl}\includegraphics[width=3in]{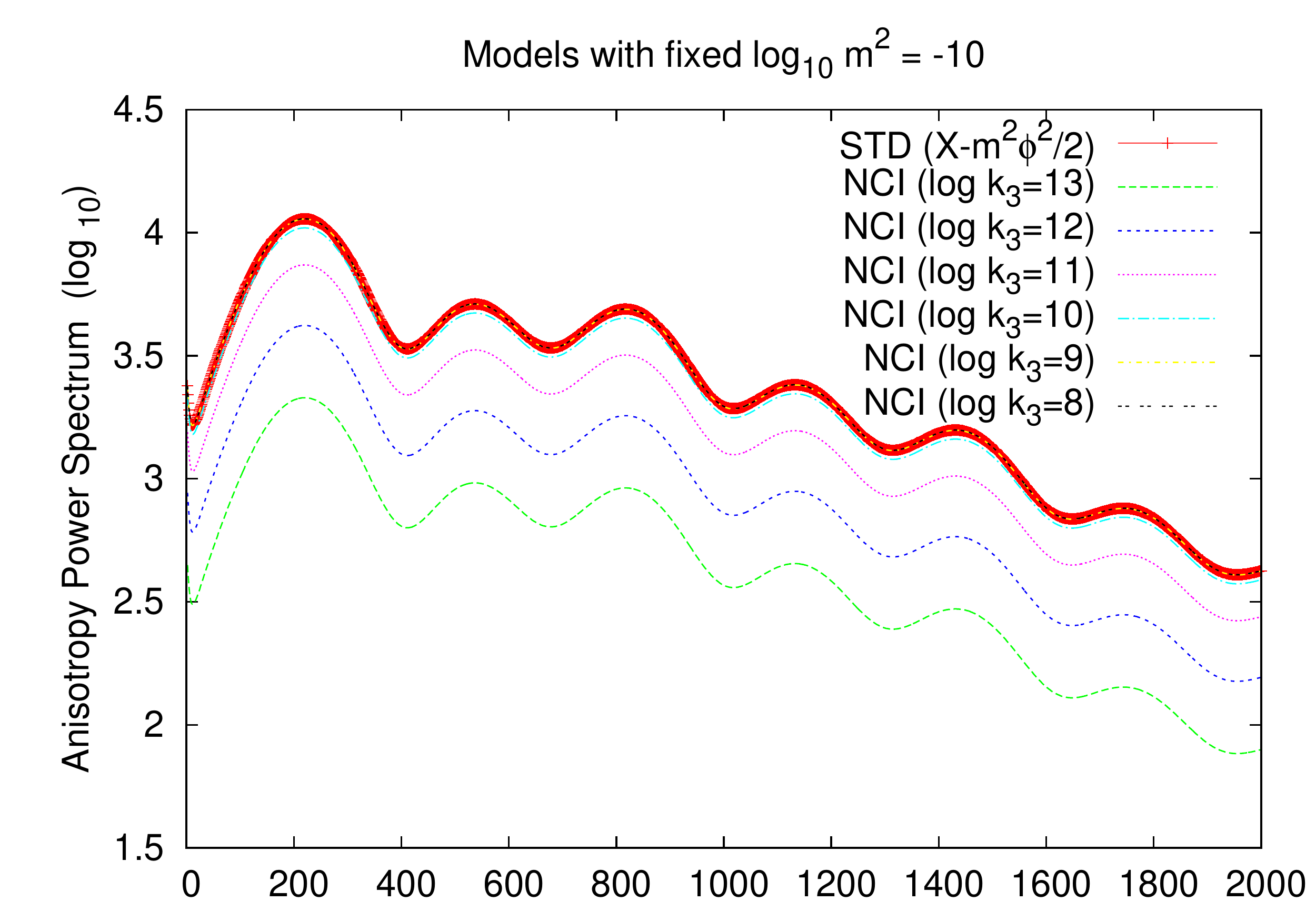}}
\subfigure[plot for quartic chaotic potential]{\label{fig:tr}\includegraphics[width=3in]{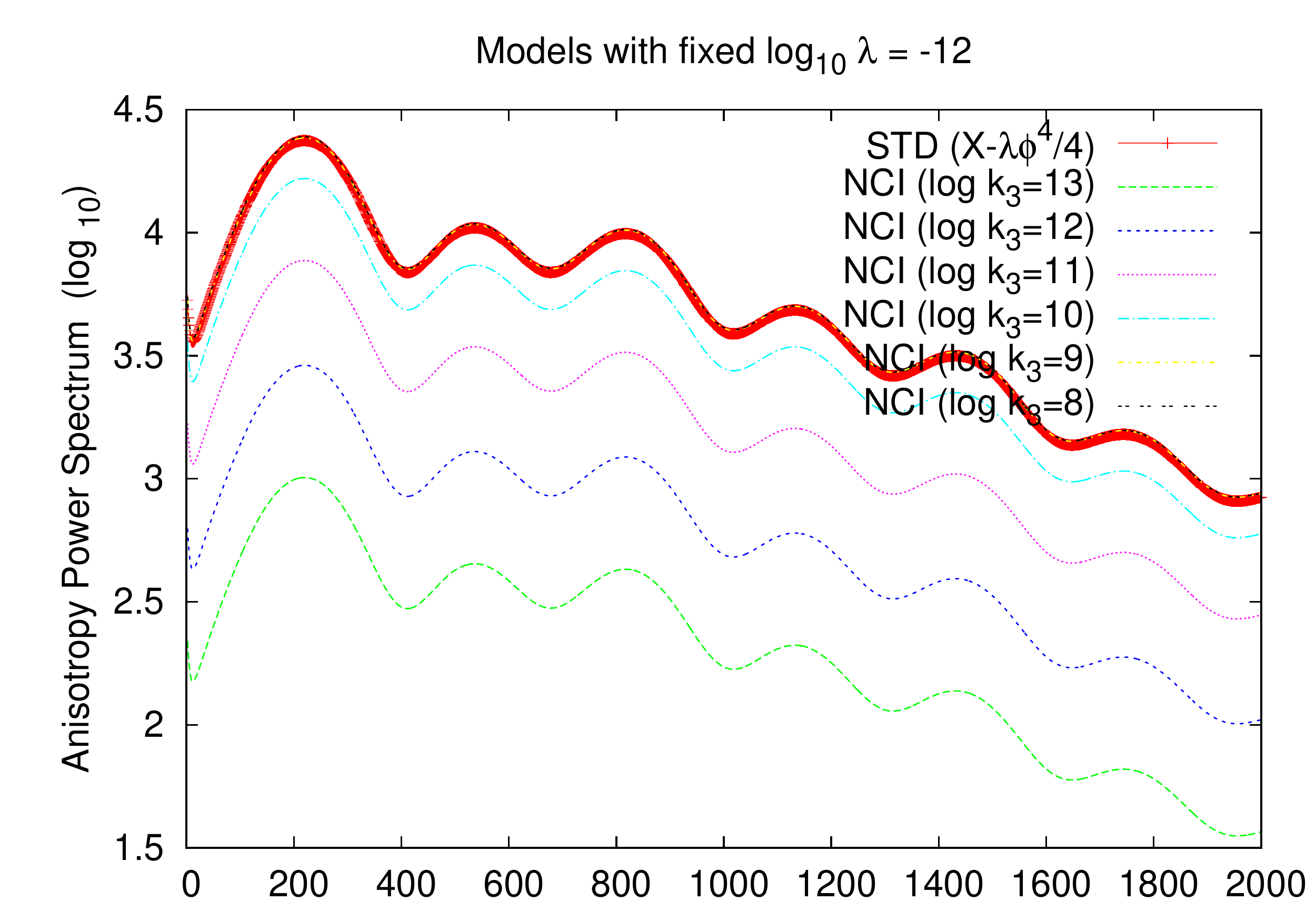}}
\caption{The effect of different values of $K_{3}$ on the final power spectrum. All numbers in the legend are base-10 logarithms, with the spectrum in units of $\mu$K$^2$. 
%In the left figure, once $\log m^{2}$ is fixed close to the value required by the observed amplitude, there is no obvious change in the final power spectrum from changing the kinetic coefficient $K_3$. In contrast, in the right figure for the quartic potential, by fixing $\log \lambda$ to the value required by observation, the final power spectrum has a significant shift due to varying $K_{3}$. This also means that for NCI models with quadratic potential, the degeneracy is strong, while for a quartic potential this degeneracy is broken more easily, at least for large $K_{3}$.
}
\label{fig:pivot-lambda}
\efg
%%%%%%%%%%%
%%%%%%%%%%%

Before we examine the results from MCMC, we look at the influence of the extra term, $K_3X^{2}$ in the kinetic function, on the final power spectrum.  The shape of the power spectrum is controlled by a combination of $A$ and $K_3$ in our considered NCI models. In the left panel of Figure \ul{\ref{fig:pivot-lambda}}, we take $\log m^{2} = {-10}$ for potential $m^{2}\phi^{2}/2$; introducing $K_3$ has negligible effect for $K_3 \lesssim 10^{10}$, above which the spectrum starts to decrease as the quadratic kinetic term becomes dominant. In the right panel we see a similar result for $\lambda \phi^4$ with $\lambda = 10^{-12}$.
Figure~\ul{\ref{fig:allcombs}} shows the spectra for some parameter values chosen so that the spectral amplitude is close to the observed value.

%%%%%%%%%%%
\bfg[t]
\centering
{\label{fig:ml}
\includegraphics[width=3in]{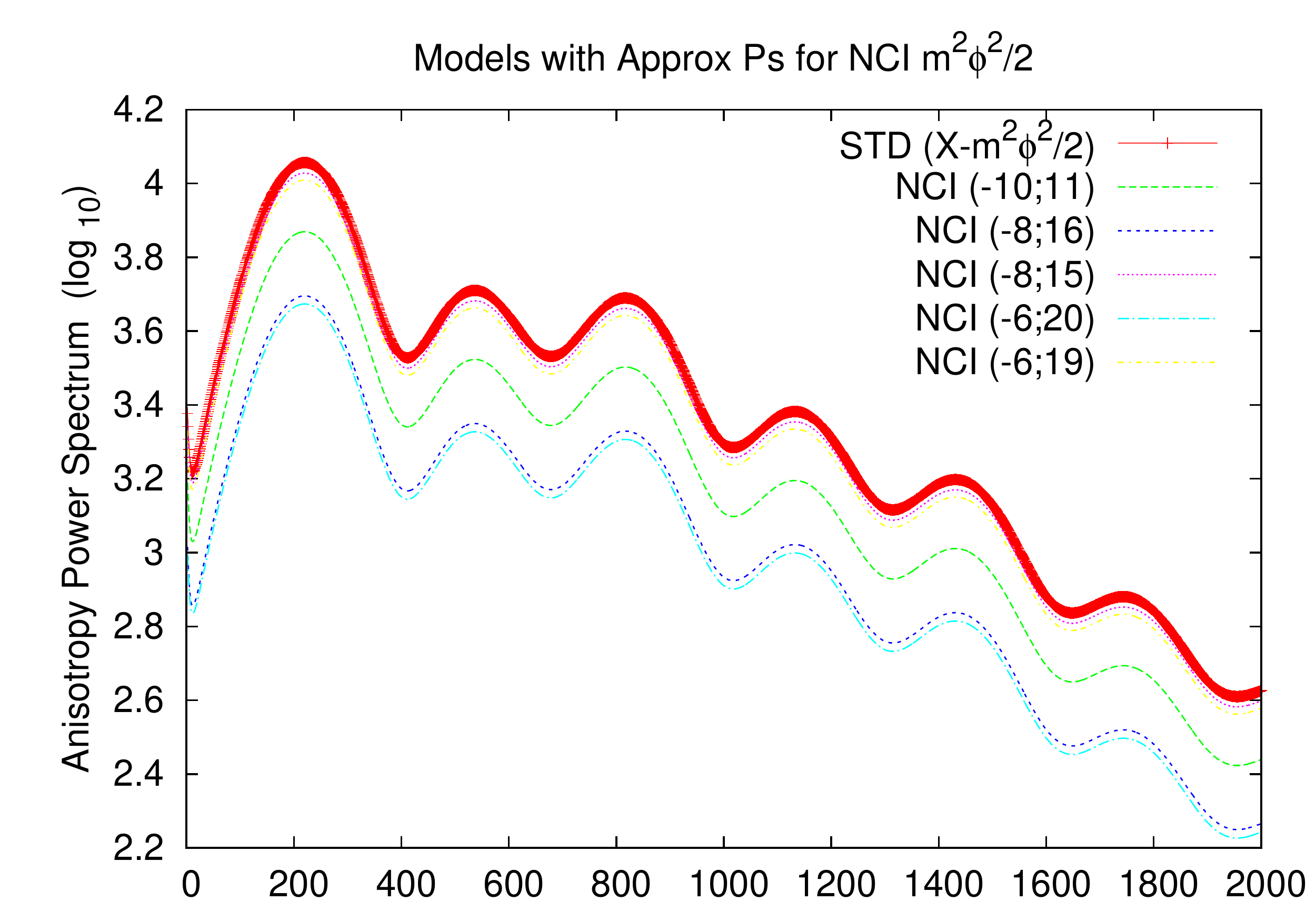}}
{\label{fig:mr}\includegraphics[width=3in]{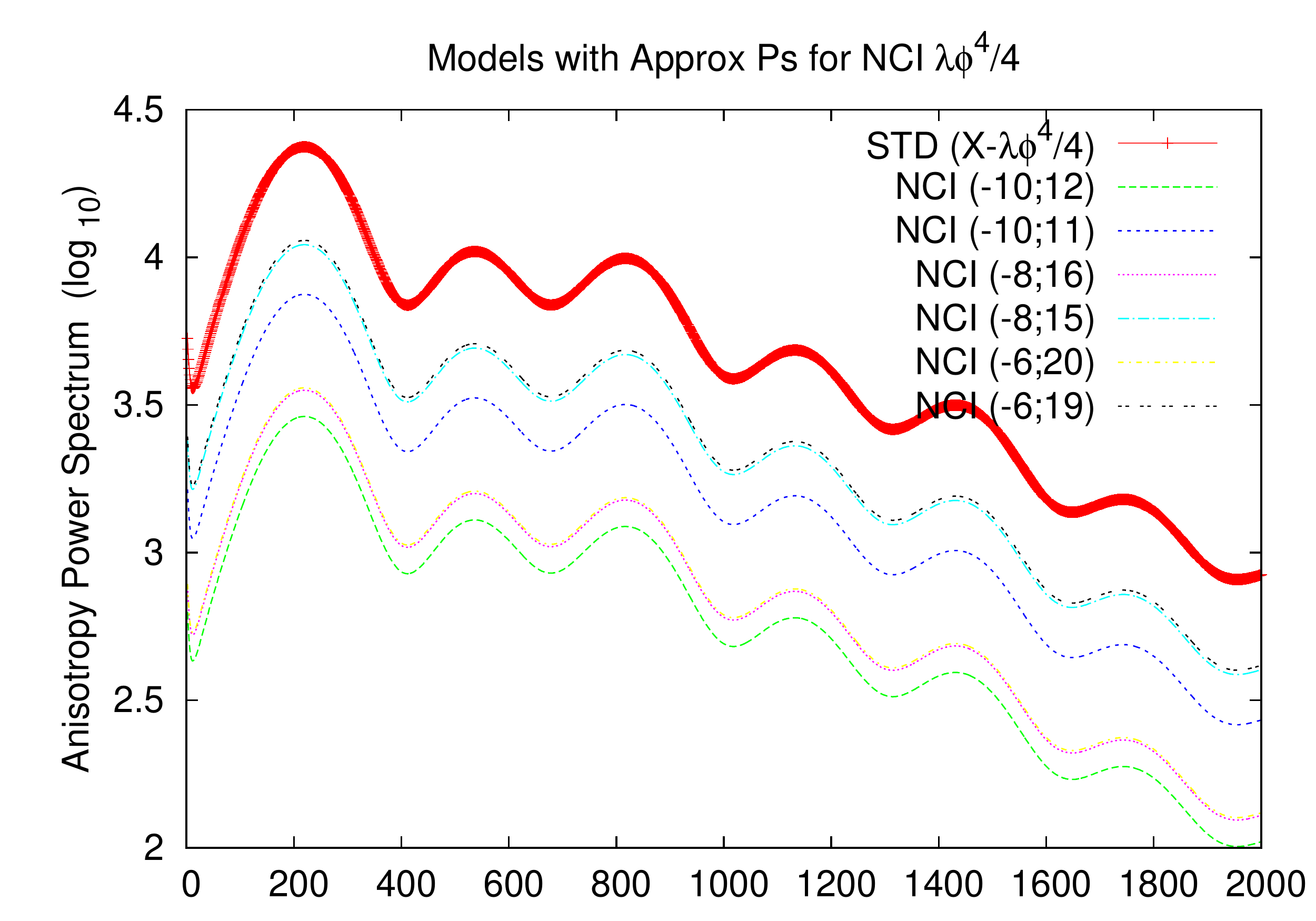}}
\caption{The power spectra for various combinations of parameters. For the NCI models, the first number in the key is the exponent of $m^2$ (left panel) or $\lambda$ (right panel), while the second number is the exponent of $K_3$.}
%The amplitudes of the scalar power anisotropy spectrum with $\log (A K_3)= {-1}$, for the quadratic (upper) and quartic (lower) potentials. The predictions for $n_{\rm s}$ and $r$ are indistinguishable; only the amplitude changes.
\label{fig:allcombs}
\efg
%%%%%%%%%%%

\subsection{Interpretation of MCMC explorations}
%Data Updated on 2nd/July;
After finishing 12 chains for each model, we have obtained a Gelman--Rubin convergence statistic $R-1 = 0.16$ 
and $0.012$ 
for the eigenvalues of the covariance matrix in models with quadratic and quartic potentials respectively. The prior ranges and maximum likelihood values are given in Table~\ref{kmc-priors}, and the posterior distributions in Figures \ul{\ref{fig:kmc-mcmc-m2p2}} and \ul{\ref{fig:kmc-mcmc-lp4}}.

%%%%%%%%%%%%%%%%%%%%%%%%%%%%%%%%%
%%%%%%%%%%%%%%%%%%%%%%%%%%%%%%%%%
\btb[b]
\centering
\bt{cccccc}
\hline\hline
Models	&\multicolumn{2}{c}{Priors} & $n_{s, {\rm ML}}$ & $r_{\rm ML}$ & $-2\ln \mathcal L_{\rm ML}$ \\ \cline{2-3}
(NCI)		&$\log A$	&$ \log K_3$\\
\hline
%Data Updated on 2nd/July;
%$(1,2;2)$ & $(-14 \,\,,\,\, -8)$ & $(0,1E14)$ & $0.965$ & $0.119$ & $7470.5$ \\
$(1,2;2)$ & $(-16 \,, -4)$ & $(0,20)$ & $0.965$ & $0.080$ & $7469.8$ \\
\hline
%$(1,2;4)$ & $(-14 \,, -10)$ & $(0,1E13)$ & $0.955$ & $0.141$ & $7473.7$ \\
$(1,2;4)$ & $(-18 \,, -4)$ & $(0,20)$ & $0.957$ & $0.115$ & $7471.8$ \\
\hline\hline
\et
\vskip .1in
\caption{Priors for the parameters, $\log A$ and $\log K_3$, and the maximum likelihood (ML) values for $n_{\rm s}$ and $r$ from WMAP7. NCI models with single-term potentials provide a red tilt, $n_{\rm s}<1$, and a detectable tensor-to-scalar ratio $r\sim0.1$.}\label{kmc-priors}
\etb
%%%%%%%%%%%%%%%%%%%%%%%%%%%%%%%%%

%%%%%%%%%%%
\bfg[t]
\centering
\includegraphics[width=3.in, height=.5\textwidth]{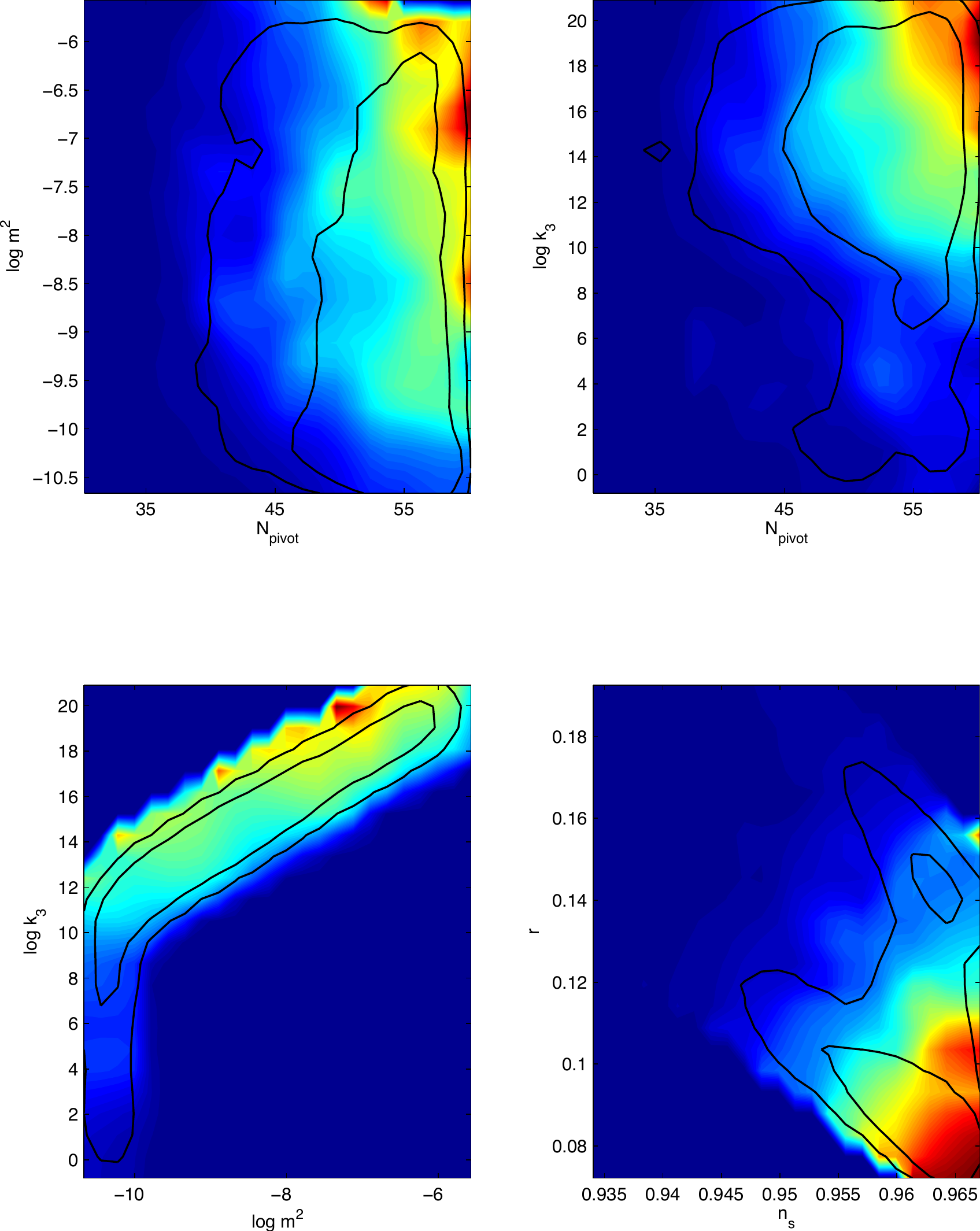}%{m2p2_jan_iii_2D_needed}%{m2p2_dec_i_2D_needed}
\hfill
\includegraphics[width=3.in, height=.5\textwidth]{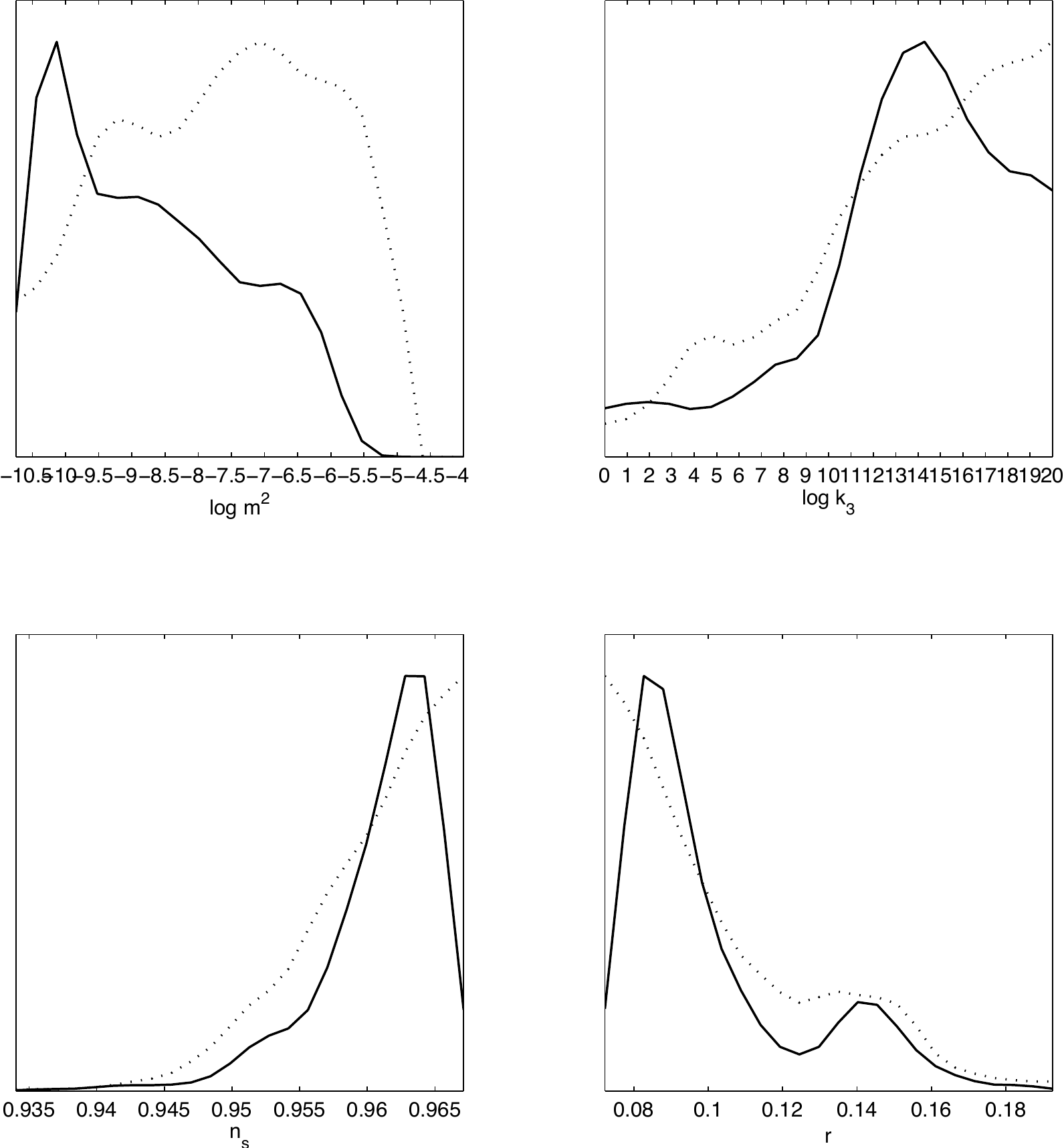}%{m2p2_jan_iii_1D_needed}%{m2p2_dec_i_1D_needed}
\caption{Parameter constraints for NCI model with quadratic potential. Left: Constraints on $\log m^2$ and $K_3$ against $N_{\rm pivot}$ for WMAP7 data. The countours are 68\% (inner) and 95\% (outer) confidence levels, while the colour scale shows the sample mean likelihood in bins. Right: one-dimensional posterior distributions for the parameters (solid) and the mean likelihoods (dashed).}
\label{fig:kmc-mcmc-m2p2}
\efg

The parameter of principal interest in each case is $K_3$. In the quadratic potential case, this parameter turns out to be completely unconstrained by the data. This is to be expected, as the quadratic potential gives acceptable observables when the kinetic term is either $X$ or $X^2$, which are the limits of small and large $K_3$. The MCMC results show that the fit to data remains acceptable right to the largest values of $K_3$ that we permit. We clearly see the two limiting behaviours of domination by either the $X$ or $X^2$ term; for example in the 2D $m^2$--$K_3$ constraint plot the former region has constant $\log m^2 \simeq -10$, while the latter has $K_3 \propto m^4$ as implied by taking constant ${\cal P}_s$ in the third column of Table \ref{abg} to obtain the observed amplitude. The bimodal likelihood of $r$ is caused by the different values of this parameter in the two regimes. We also see a very mildly enhanced likelihood in the transition regime $K_3 \simeq 10^{12}$, but all values of $K_3$ are acceptable  Perhaps surprisingly, then, present data can say nothing about the amplitude of a quadratic kinetic term added to the normal canonical one for this potential, and as this is a potential known to fit the data well in the canonical case we can conclude that more generally a quadratic correction term cannot be constrained directly from data.

The quartic potential case shown in Figure \ul{\ref{fig:kmc-mcmc-lp4}} is more interesting. In the standard cosmology, $K_3 \rightarrow 0$, the quartic potential is quite disfavoured by WMAP7. As we would anticipate from the slow-roll results of Section \ref{sec:sr}, the situation for this potential actually improves in the limit of large $K_3$, as both $n_{\rm s}$ and $r$ move towards the scale-invariant values. Once more the MCMC results show that the fit remains good as the quadratic kinetic term becomes dominant and we find no observational upper limit on $K_3$ for this potential. This time the distribution for $r$ is unimodal, as only the $X^2$ domination regime contributes to the posterior. There is a plausible non-zero lower limit on $K_3$, though the numerical value of such a limit will be quite prior-dependent. Hence, incorporation of an $X^2$ term is a method of salvaging the quartic model, though the large value of $K_3$, in Planck units, that is required is unattractive. 

%%%%%%%%%%%
\bfg[t]
\centering
\includegraphics[width=3in, height=.5\textwidth]{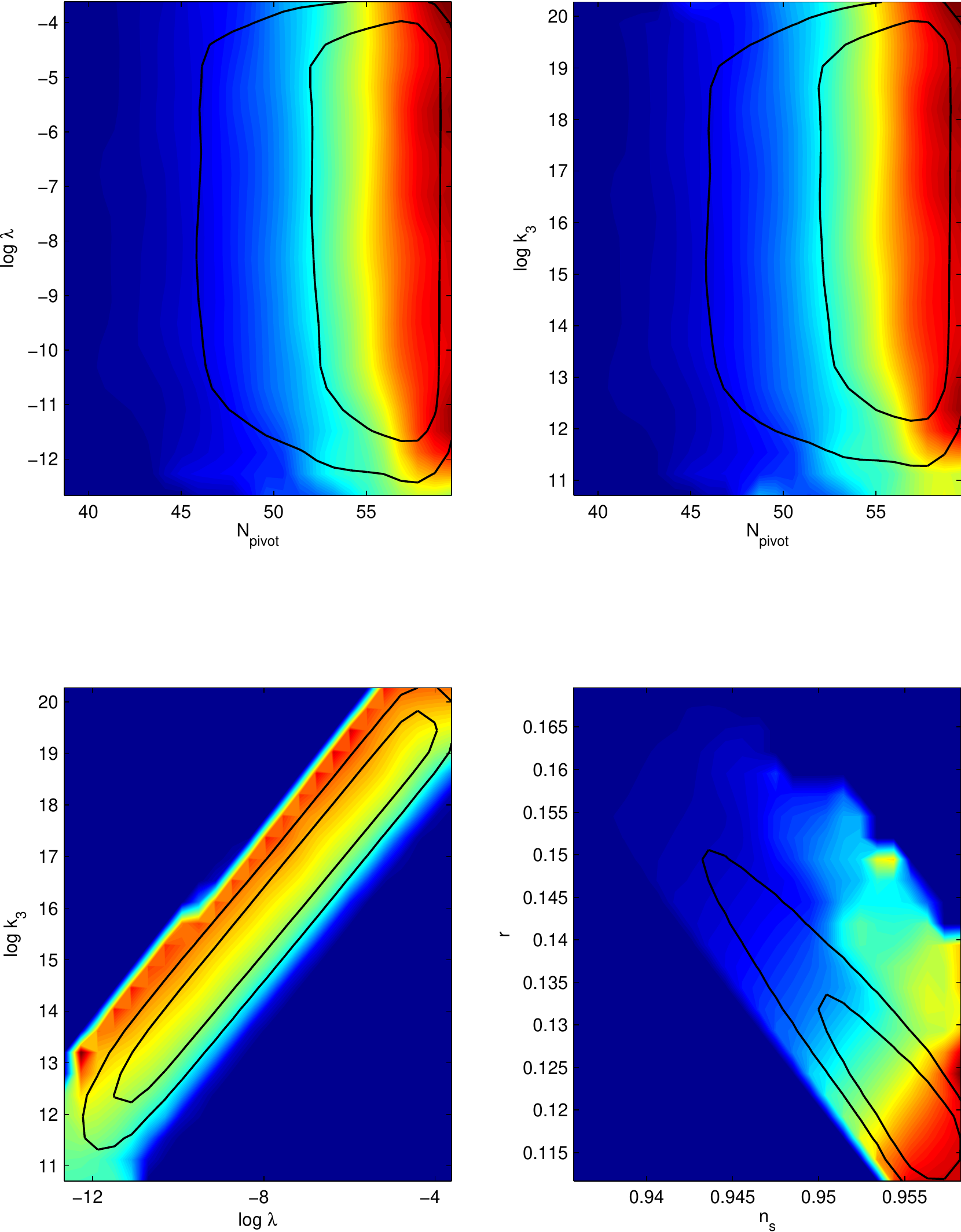}%{lp4_jan3_2D_needed}%{lp4_jan_iii_2D_needed}%
\hfill
\includegraphics[width=3in, height=.5\textwidth]{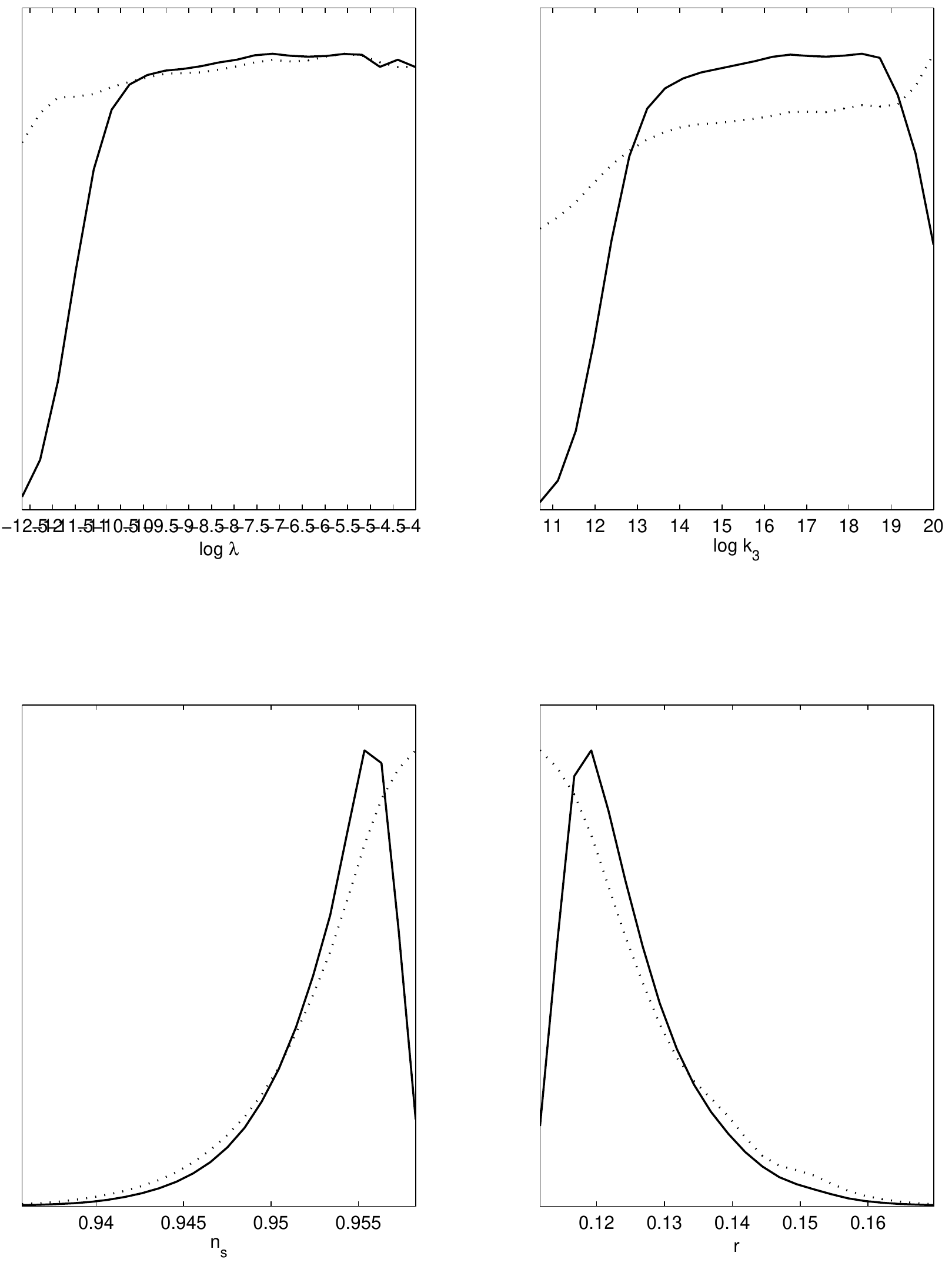}%{lp4_jan3_1D_needed}%{lp4_dec_i_1D_needed}
\caption{Parameter constraints for NCI model with single term quartic potential. This figure uses the same convention as in figure~\ul{\ref{fig:kmc-mcmc-m2p2}}.}
\label{fig:kmc-mcmc-lp4}
\efg

\section{Conclusions}

In this paper we introduced a numerical solver \textbf{Kinetic Module Companions (KMC)}, an extension to ModeCode for a class of non-canonical inflation (NCI) models. In this article we have used our code to investigate some simple non-canonical models, which have up to two terms in the kinetic energy and a monomial potential, in order to test the validity of the code and provide some initial scientific results. We found that these models are well able to fit current data, including in the quartic potential case provided the quadratic kinetic term dominates. This is compatible with slow-roll results we obtained in the case of a single kinetic term of power-law form.

As a specific application of the code, we studied the introduction of a quadratic correction $K_3 X^2$ to the normal canonical kinetic term $X$, with the goal of constraining the coefficient $K_3$. In practice, however, $K_3$ turns out to be unconstrained by data, and indeed the inclusion of a large quadratic term can even improve the fit to WMAP7 data, for instance for a quartic potential which, in the canonical case, is under severe pressure from observations. Accordingly present data allow no leverage whatsoever on radical deviations from the canonical case. In future work we plan a much more comprehensive investigation of possible kinetic and potential forms. 

A longer-term objective in this area may be to extend ModeCode to yet more complex forms of single scalar-field action, such as the Galileon \cite{galileon} or indeed the Horndeski action \cite{Horn} which is the most general scalar--tensor theory yielding second-order equations of motion. However the many functional degrees of freedom of such actions will no doubt lead to considerable degeneracies given the relatively limited amount of observational information available, which essentially amounts to only a couple of numbers at present. Hence, as we have found here even for the simplest separable K-inflation case, one is likely to need considerable guidance from theory as well as from observations in assessing whether the most general paradigms are useful.

\vskip 6pt

{\em Noted added:} Shortly after we put this article on the arXiv, an independent paper \cite{UST} was submitted which contains results that partially overlap with ours.

\acknowledgments
S.L.\ was supported by a Sussex International Research Scholarship, and A.R.L.\ by the Science and Technology Facilities Council [grant numbers ST/F002858/1 and ST/I000976/1] and a Royal Society--Wolfson Research Merit Award. We thank the authors of ModeCode for making their code public, and two of them, Richard Easther and Hiranya Peiris, for discussions in relation to the work reported here. We also thank Antony Lewis, David Seery and Wessel Valkenburg for useful discussions. We acknowledge use of the Apollo HPC cluster at the University of Sussex and the Sciama High Performance Compute (HPC) cluster which is supported by the ICG, SEPnet and the University of Portsmouth.

%%%%%%%%%%%%%%%%%%%%%%%%%%%%%%%%%%%%%%%%%%%%

\end{document}